\documentclass{Interspeech}

\interspeechcameraready

\title{Fifteen Years of Child-Centered Long-Form Recordings:\\ Promises, Resources, and Remaining Challenges to Validity}

\author[affiliation={1}]{Loann}{Peurey}
\author[affiliation={2}]{Marvin}{Lavechin}
\author[affiliation={1}]{Tarek}{Kunze}
\author[affiliation={1}]{Manel}{Khentout}
\author[affiliation={3,1}]{Lucas}{Gautheron*}
\author[affiliation={1}]{Emmanuel}{Dupoux}
\author[affiliation={1}]{Alejandrina}{Cristia*}

\affiliation{LSCP}{DEC, ENS, EHESS, CNRS, PSL University}{France}
\affiliation{BabyDevLab}{University of East London}{UK}
\affiliation{IZWT}{University of Wuppertal}{Germany}
\email{*lucas.gautheron@gmail.com, *alejandrina.cristia@ens.psl.eu}


\usepackage{pgfplots}
\pgfplotsset{compat=1.18}
\usepackage{booktabs}
\usepackage{amsmath}
\usepackage{bm}
\usepackage{subcaption}
\usepackage{graphicx}
\usepackage{placeins}
\usepackage{float}

\usepackage{enotez}
\DeclareInstance{enotez-list}{interspeech}{paragraph}{
  heading={\raggedright#1},
  notes-sep=0pt,
  number=\makebox[0pt][r]{#1.\ }\ignorespaces,
 }

\begin{document}
\maketitle

\begin{abstract}
   Audio-recordings collected with a child-worn device are a fundamental tool in child language research. Long-form recordings collected over whole days promise to capture children's input and production with minimal observer bias, and therefore high validity. The sheer volume of resulting data necessitates automated analysis to extract relevant metrics for researchers and clinicians. This paper summarizes collective knowledge on this technique, providing entry points to existing resources. We also highlight various sources of error that threaten the accuracy of automated annotations and the interpretation of resulting metrics. To address this, we propose potential troubleshooting metrics to help users assess data quality. While a fully automated quality control system is not feasible, we outline practical strategies for researchers to improve data collection and contextualize their analyses.

\end{abstract}

\section{Introduction}
Audio-recordings collected with a child-worn device are a fundamental tool in research on child language \cite{macwhinney1998childes}. The last decade has seen increasing use of long-form recordings, collected as children wear a device typically over a whole day, to capture what children hear and what they say \cite[to cite just a few]{cychosz2025bursty,dupas2023informing,bergelson2019day,bergelson2023everyday,scaff2024characterization}. In the context of the Special Session ``Challenges in Speech Data Collection, Curation, and Annotation'', this paper seeks to provide an entry point to this burgeoning literature (Sections 2-4), as well as make one novel contribution by clarifying extant challenges to the quality of automated annotations as well as the interpretation of resulting metrics (Section 5).

\section{Definition and Key Uses of Long-Form Recording Data}
Compared to short-form recordings, which often take place during a specific activity, long-form recordings aim to capture speech behavior ``in the wild'': Families are asked to go about their normal day as much as possible. The hope is that the participation burden is lowered for families in this way, and that observer effects (whereby the behavior of those recorded is affected by the feeling of being ``watched'') are minimized. Some research shows that families do behave differently in long-form recordings as opposed to shorter ones \cite{bergelson2019day}, although of course there is no way to ensure that there is absolutely zero effect of being observed.

\begin{figure}
    \centering
    \includegraphics[width=\linewidth]{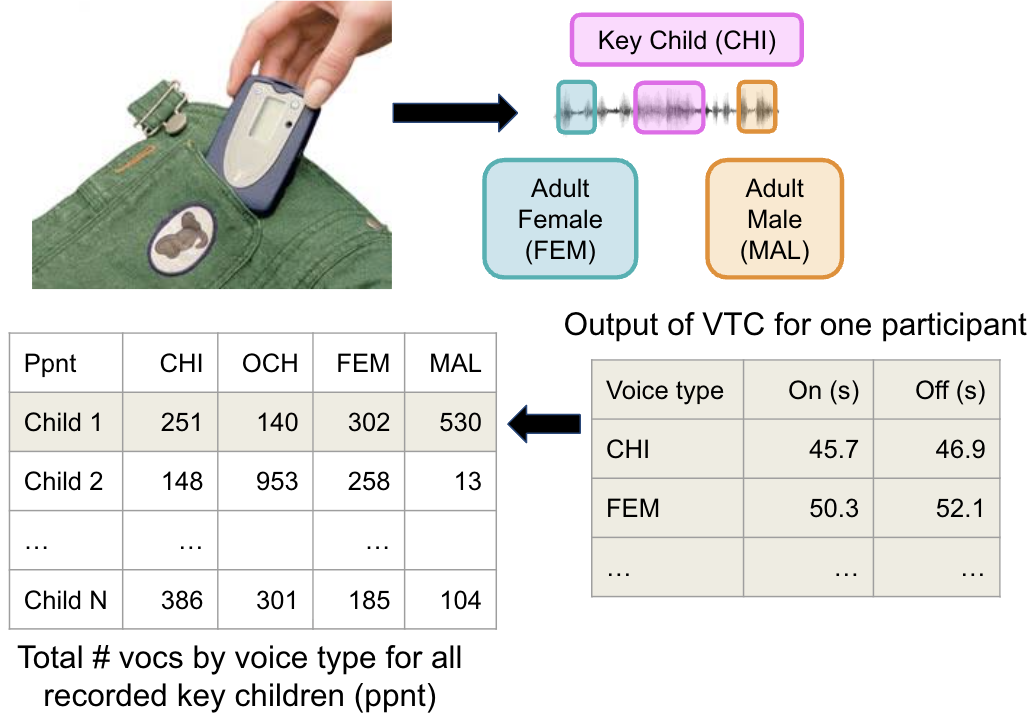}
    \caption{Long-form recordings, gathered with a small wearable, are submitted to a voice type classifier to extract potentially  metrics like the number of vocalizations by different talker types: the child wearing the device (CHI), other children (OCH), and adult females (FEM) and males (MAL).}
    \label{fig:setup}
\end{figure}

Each child's speech input and output is thus recorded in sessions that are often 8-16 hours long. The resulting volume of audio data necessitates automated processing, often with specialized algorithms, since the data are too challenging for off-the-shelf solutions. Due to space limitations, we focus here on information that results from one type of such algorithm called voice-type classifiers, whereby the audio is automatically parsed into silence as opposed to four voice types: the key child wearing the device (CHI), male and female adult (MAL and FEM), and other children around the key child (OCH). This results in estimates of the number of vocalizations in children's input and output (Figure \ref{fig:setup}).

The technique of long-form recordings is poised to revolutionize multiple fields. In psycholinguistics, one study leveraged its big-data potential by using over 40k hours of audio from 1,001 children to assess possible predictors of children's spontaneous vocalizations \cite{bergelson2023everyday}, for instance finding a significant association between how much children vocalized and the amount of adult speech, but not the child's gender. Another study explored whether self-supervised learning models develop phonological attunement \cite{lavechin2024modeling}, concluding that to become specialized in the ambient phonology represented in long-form data, model learners may need biases like distinguishing speech from non-speech.

Long-form recordings also show great promise in applied research. For instance, they may be useful in geographic areas where speech-language pathologists are scarce and automated tracking could save their valuable time \cite{hamrick2023semi, cychosz2022using}. Public health specialists and economists increasingly use long-form recordings to assess childhood interventions. One study found that families exposed to a 3-minute video on the importance of talking to children showed increased adult speech, measured by automated analysis of 9 hours of audio from 449 families \cite{dupas2023informing}.

\section{Extant Resources for Facilitating Data Collection, Human Annotation and Sharing}

A range of resources exist to help researchers navigate the collection, annotation, and sharing of long-form recordings. In addition to introductory papers \cite{casillas2019step,casillas2024daylong}, there exist also some video tutorials\endnote{\url{https://youtube.com/playlist?list=PLExuQICGVy3gMQllaZ5OLDOQ1KDMuhoMr}} and structured documentation\endnote{doi:10.5281/zenodo.6685828}. A summer school with spin-offs in several countries is planned for  2025.\endnote{\url{https://lfraz2025.sciencesconf.org/}} Among the most useful resources, the ACLEW annotation scheme provides a framework for human coding child-centered recordings that combines flexibility with consistency across studies, facilitating cross-laboratory comparisons \cite{soderstrom2021developing}. 

Resources exist helping researchers navigate the complex ethical issues involved in such recordings \cite{cychosz2020longform}. Given that long-form recordings capture children’s everyday lives in an unfiltered manner, they cannot be made publicly available (without thorough vetting). A handful of full-length audios have undergone vetting and have been rendered fully public through HomeBank\endnote{\url{homebank.talkbank.org}} (the TalkBank section dedicated to long-form recordings, \cite{vandam2016homebank}). However, most datasets are private, including BabyTrain, which was used for developing what is currently the only open-source and well-documented voice type classifier (called VTC) \cite{vtc}. 

The community of long-form recordings using child wearables is open and helpful. The network DARCLE.org has a mailing list and a monthly seminar, providing a unique space for knowledge exchange and discussion. Early Career Researchers benefit from a supportive monthly peer group meeting, to which leading scholars are sometimes invited, allowing students and post-docs to strengthen their networks and get crucial feedback on their projects.

\section{Accuracy of Automated Voice Type Classifiers for Long-Form Recordings}

In many cases, researchers using this technique must rely on automated metrics rather than human annotations. The LENA software system \cite{gilkerson2020guide, gilkerson2008lena, xu2008reliability}, a widely used proprietary tool for long-form recordings, has been the focus of extensive accuracy evaluations. A 2020 review of 33 benchmarking studies raised three major concerns \cite{cristia2020accuracy}: (1) few studies assessed all four key voice types (key child, other children, female adults, male adults), (2) average reported precision (59\%) and recall (64\%) were subpar, and (3) performance varied widely across studies, with recall ranging from 11\% to 80\%, limiting generalizability even when using the LENA hardware-software suite.

An open-source alternative, the above-mentioned VTC \cite{vtc}, has been less studied but shows similar variability. In an unpublished analysis, Bergelson et al. compared human-automated correlations across seven datasets. Median correlations were .75 for VTC (.77 for LENA), but variability was high: .4–.81 for VTC and .51–.89 for LENA, reinforcing concerns about reliability across datasets. A similar model was recently released \cite{li2024enhancing}, but remains untested on datasets and languages outside those reported in the original study.

\section{Challenges to the Interpretation of Automated Metrics, and Potential Solutions}

Having set the background on this technique for Special Session attendees, in this section we move on to the major novel contribution of this paper, which is providing a framework for potentially understanding (and eventually correcting) issues that complexify the interpretation of automated metrics. Both researchers that are collecting data in a new environment, language, or culture and/or at scale; and those who aim to re-use archived data will wonder: (Q1) Can I trust the automated output of a system for each and every session and participant? (Q2) Are there ways in which I can  troubleshoot for issues affecting the interpretation of the resulting metrics? 

\subsection{\label{section:variation}A Systematization of Sources of Variation in Algorithm Performance and Resulting Metrics}

Conceptually speaking, sources of variation in algorithm performance and the resulting metrics across corpora and recordings (Q1 above) arise from five main sources (see Table \ref{tab:problem_classification}): (a) the hardware used to gather the data, (b) the way the device is placed and operated by the participants,  (c) the circumstances under which the data is recorded, (d) the social environment of the child (e.g. the family structure), and (e) the children themselves. 

The vast majority of researchers employ one specific \textbf{hardware}, conceived and commercialized by the LENA Foundation\endnote{\url{www.lena.org}}. As an increasing number of researchers turn away from LENA products, recording devices vary widely, from high-quality, high-bit-rate models (e.g., Olympus) with flat frequency response to low-bit-rate USB devices with unknown specifications. Researchers who aimed to collect data from hundreds of families have turned to USB devices, sometimes sourced from different sellers, because the cost of data collection was prohibitive otherwise. 

Beyond hardware differences, protocol deviations introduce \textbf{operating errors}. Some issues are easy to detect, such as families turning devices on and off instead of recording continuously, since this is reflected in shorter files. Others are more subtle — while devices are meant to be worn on the child’s chest, some families might leave them on a table or even in a drawer.

And there are other sources of between-children variation that require us to take the output of automated analyses with a grain of salt through no fault of the technique, since variation emerges from differences in e.g. acoustic properties of the child environment (\textbf{setting}), variation across \textbf{families} or across \textbf{children} themselves. Perhaps the family routinely leaves the TV on; asking them to turn it off would not represent the children's audio environment, but leaving it on may lead to confusion between live and pre-recorded voices.
Similarly, if the child has siblings, accuracy for the CHI and OCH categories may be lower if the algorithm confuses the siblings with the child wearing the recorder.

\subsection{\label{section:attempts}Initial Attempts at Handling Variability in Performance and Metrics}

As mentioned above, it would be ideal to have ways to troubleshoot and diagnose issues (Q2). In this section, we show that, in fact, hypothesized negative effects in Table \ref{tab:problem_classification} are not always verified, leading us to the conclusion that further methodological work is needed. 

\begin{table*}[h]
    \centering
    \caption{Hypothetical classification of challenges to metrics emerging from automated analyses, in some portion of the long-form recording or the whole recording, as indicated by Scope: W=whole recording, H=hours, M=minutes. . VTC stands for Voice Type Classifier (VTC), which diarizes the audio into the following voice types: CHI$=$key child --wearing the device, FEM$=$female adults, MAL$=$male adults, OCH$=$other children). Its performance may be reduced for all speakers ($\forall$) or just for some of them; resulting in mis-estimation of derived metrics like number of vocalizations (nb vocs). If the  hypothesized effect was studied in this paper, we indicate in Fig where results can be found.  * requires norming (e.g. by child age, time of day, etc.), much of which is not available.}
    \begin{tabular}{p{5cm} p{1cm} p{5cm} p{5cm} }
        \toprule
        \textbf{Problem } & \textbf{Scope} & \textbf{Hypothesized Negative Effect} & \textbf{Potential Troubleshooting Indicator} \\
        \midrule
\hline	&	& \bf{Source: Hardware}	&	\\
Recorder is cheap 	&W	&$\downarrow$ audio quality, $\downarrow$ VTC perf. $\forall$ (\bf{Fig.  \ref{fig:device}})	&  \textit{Not necessary}	\\
Recorder has AI denoising	&W	&Audio data does not fit trained models, $\downarrow$ perf. VTC $\forall$	& ?	\\
Recorder loses samples	&M	&Discontinous audio context	& ?	\\
Recorder not tight against the child	&W	&Increased friction noise	&$\downarrow$ SNR	\\
\hline	&	& \bf{Source: Operating error}	&	\\
Metadata on start time incorrect causing daytime \& nighttime confusion	&H/W	&Input nb vocs mis-estimated, loss of contextual information	& $\downarrow$ nb vocs  $\forall$*	\\
Recorder not on child	&M/H/W	& $\downarrow$ perf. VTC CHI $\rightarrow$ CHI missed	&$\downarrow$ CHI nb vocs*, $\downarrow$ SNR for CHI	\\
\hline	&	& \bf{Source: Setting}	&	\\
Room size \& quality leading to $\uparrow$ reverberation 	&M/H	&$\downarrow$ VTC perf. $\forall$	&$\uparrow$ C50 (\bf{Fig.  \ref{fig:c50}})	\\
$\uparrow$ Background noise 	&M/H/W	&$\downarrow$ VTC perf. $\forall$	&$\downarrow$ SNR (\bf{Fig.  \ref{fig:snr}})	\\
Outdoors with $\uparrow$ environmental noises 	&M/H	&$\downarrow$ VTC perf. $\forall$ (\bf{Fig.  \ref{fig:rural}})	&$\downarrow$ SNR \& $\downarrow$ C50 	\\
TV/radio on	&M/H/W	&More background noise	&$\downarrow$ SNR \&/or increase in false alarm	\\
\hline	&	& \bf{Source: Family}	&	\\
Infant's primary caregivers not FEM	&W	&Input nb vocs mis-estimated because algorithms have $\uparrow$ perf. for FEM than MAL/OCH	&$\downarrow$ FEM nb vocs*, $\uparrow$ nb vocs MAL/OCH*	\\
Child is carried \&/or swaddled a significant portion of the day	&W	&CHI missed because of mic obstruction	&$\downarrow$ CHI nb vocs*, $\downarrow$ SNR $\forall$	\\
More crowded household  	&M/H/W	& More overlapping speech \& background noise	&$\uparrow$ nb vocs FEM/MAL/OCH*, $\uparrow$ SNR 	\\
Other children present	&M/H/W	&More confusion OCH/CHI	&$\uparrow$ CHI nb vocs*	\\
\hline	&	& \bf{Source: Child}	&	\\
Child's voice departs from training data	&W	&CHI missed	&$\downarrow$ CHI nb vocs*	\\
Child sleeps more, less, or differently	&W	&(Variable)	&?	\\
        \bottomrule
    \end{tabular}
    \label{tab:problem_classification}
\end{table*}
\FloatBarrier

For the analyses below, we were able to rely on private long-form data shared with us through agreements. Due to space limitations, we provide readers with the dataset name and the reference where more information on the datasets can be found: bergelson (dataset: \cite{bergelson2019day}), lucid (dataset: \cite{Rowland_Durrant_Peter_Bidgood_Pine_Jago_2024}, warlaumont (dataset: \cite{warlaumont_corpus}), winnipeg (dataset: \cite{mcdivitt_corpus}; for all of the previous corpora, annotations are described in \cite{soderstrom2021developing}); cougar (dataset: \cite{vandam_cougar}; annotations: \cite{vandam_5min}); tsimane2017 (dataset and human annotation: \cite{scaff2024characterization}). 

\subsubsection{Algorithm performance varies across hardware in non-obvious ways}
\begin{figure}[h]
    \centering
    \begin{subfigure}[t]{0.49\linewidth}
        \centering
        \includegraphics[width=\linewidth]{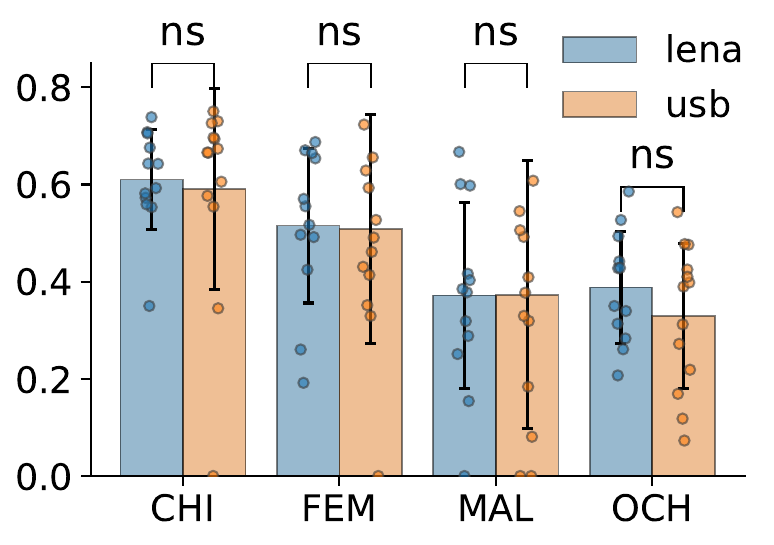}
        \caption{F-score by class and hardware.}
        \label{fig:device}
    \end{subfigure}
    \hfill
    \begin{subfigure}[t]{0.49\linewidth}
        \centering
        \includegraphics[width=\linewidth]{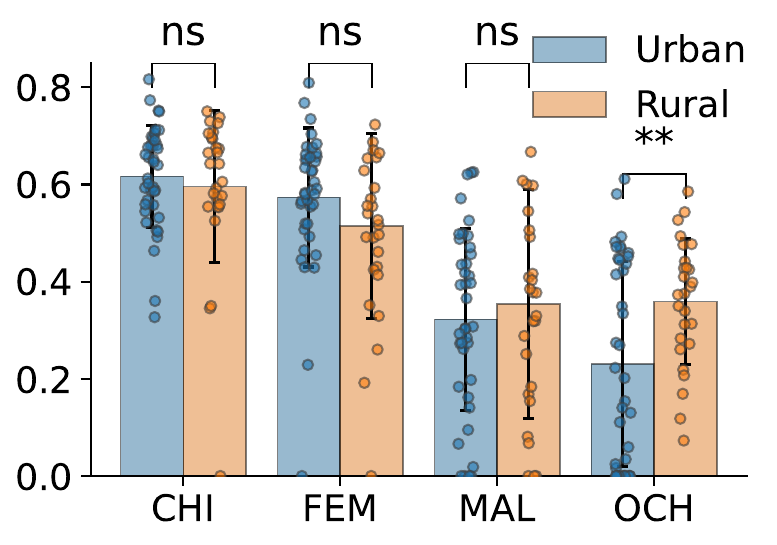}
        \caption{F-score by class and community type.}
        \label{fig:rural}
    \end{subfigure}
    \caption{Surprising (non-)effects of hardware and setting (urban versus rural). Each point indicates one child. Data for the left panel: tsimane2017. Data for the right panel: tsimane2017 (rural, LENA and USB); bergelson, lucid, warlaumont, and winnipeg (all urban, LENA only).}
    \label{fig:device_rural}
\end{figure}

An unpublished analysis\endnote{\url{https://gin.g-node.org/LAAC-LSCP/longform-hardware-audio-test}} using the exact same set-up (laboratory re-recording of an audio containing 5 minutes of short-form and 5-minutes of long-form) revealed wide variation in VTC performance across two USB devices that were identical in their appearance, with an unweighted average F-score of 45\% for one currently costing 21\$ in Amazon and 68\% for another currently costing 14\$. In the same laboratory experiment, the F-scores from the LENA software applied on the audio collected with the LENA hardware (which costs 200-400\$) achieved an F-score of 67\%. The fact that the advantage does not necessarily go to the pricier option also becomes obvious in an analysis of the tsimane2017 dataset, which includes data collected using LENA and USBs, as obvious in Figure \ref{fig:device}. Together, these results suggest that there may not be an easy solution, whereby we discard any data not collected with LENA; but also that we need to be mindful of the diverse performance that can be achieved with seemingly similar devices.

\subsubsection{Algorithm performance varies across Setting/Family/Child in non-obvious ways}
Table 1 also suggests negative effects on algorithm performance, among other negative effects, across various sources beyond hardware. We checked this expectation through an admittedly indirect route: We compared datasets collected in urban, industrialized settings versus rural, small-scale communities. Algorithms, including VTC, should perform more poorly in the latter than the former for several reasons. To begin with setting differences, there should be a great deal of environmental background noise in the rural datasets. For instance, without solid walls isolating them from nature, Tsimane' children's audio environment is characterized by vocalizations by birds, pigs, and crickets punctuating the hum of soft adult conversations carried out in the background.  In terms of family differences, families were much larger, with more siblings and other children around in the rural than urban datasets. Turning to the Child source, although VTC was intentionally trained on a linguistically diverse dataset, it was nonetheless the case that a majority of the training data came from city-dwelling, nuclear families with no or few siblings present in the home during the recordings, which are a better fit for our urban dataset. 

Once again, we were humbled by the fact that results contradicted our predictions. VTC performance is not significantly worse for all four classes in urban versus rural datasets  (see Figure \ref{fig:rural}, and supplemental information of \cite{bergelson2023everyday} for a similar result for LENA). In addition, we found no significant differences for the key-child (CHI) category in F-scores for children who, according to metadata, had siblings (average F-score=60\%, N=18) versus those who did not (average F-score=62\%, N=20; p=.52). In fact, performance for the other child (OCH) label was significantly better for children with (average F-score=40\%, N=18) versus without siblings (average F-score=10\%, N=20; p$<$.01), due to the fact that this label had much higher precision in the former than the latter case (with similar recall).

\subsubsection{Algorithm performance varies across audio conditions in predicted ways}
\begin{figure}[h]
    \centering
    \begin{subfigure}[b]{0.49\linewidth}
        \centering
        \includegraphics[width=\linewidth]{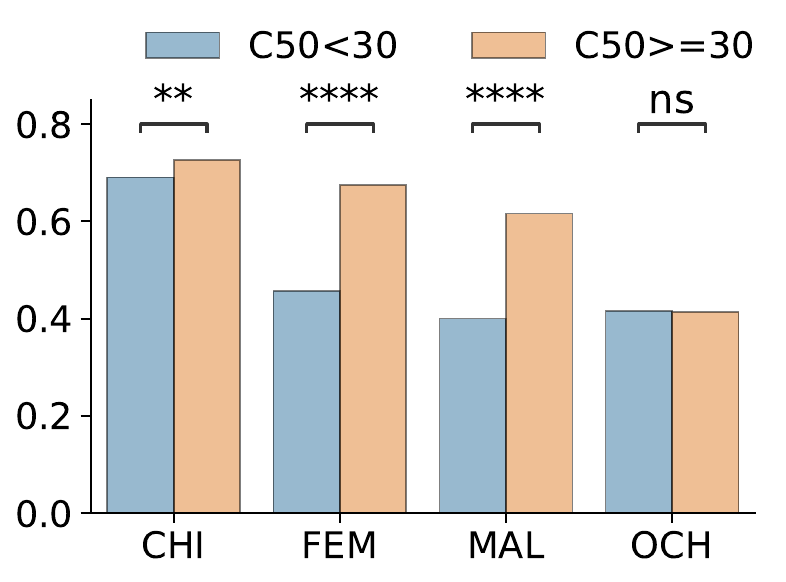}
        \caption{Correct detection by class and automated reverb estimates.}
        \label{fig:c50}
    \end{subfigure}
    \hfill
    \begin{subfigure}[b]{0.49\linewidth}
        \centering
        \includegraphics[width=\linewidth]{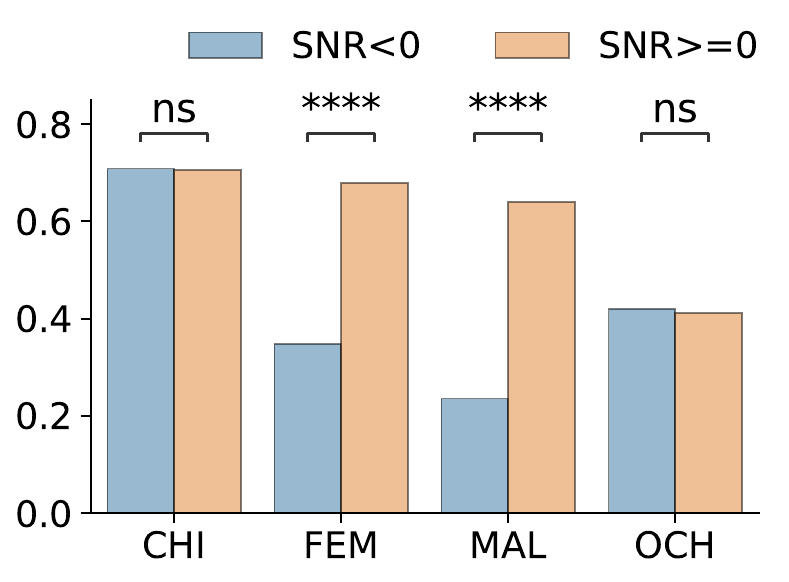}
        \caption{Correct detection by class and automated SNR estimates.}
        \label{fig:snr}
    \end{subfigure}
    \caption{VTC performance is significantly higher in the FEM and MAL categories for audio sections with high C50 (low estimated reverberation) and low (estimated) SNR. Bars indicate averages across all annotated 1-second windows (the Y variable is 1 if the speaker was correctly detected, 0 otherwise). Data: tsimane2017, bergelson, warlaumont, and winnipeg.}
    \label{fig:c50_snr_f}
\end{figure}

Even though results did not validate our hypothesized negative effects so far, we have gone ahead and assessed the possibility that at least some of the troubleshooting indicators in Table 1 may indeed predict algorithm performance in useful ways. Specifically, we focused on two potentially useful automated estimates returned by the ``Brouhaha'' \cite{brouhaha} system, C50 as an automated estimate for reverberation, and SNR for background noise, which we thought would be helpful when troubleshooting problems emerging from setting differences. Figure \ref{fig:c50_snr_f} shows that correct detection for female and male adult voice classes is lower when more reverberation and lower signal-to-noise are automatically detected, with reverb also affecting key child correct detection. 
This is promissing, but it would remain to be shown that more accurate measures of e.g. number of vocalizations in the input may ensue if sections of the audio that are high in reverberation and/or low in SNR are altogether excluded from analyses. 

\subsubsection{Number of vocalizations as potential troubleshooting indicators}
We originally thought that it would be trivial to generate distributions of e.g. number of vocalizations by different voice types, allowing us to show that researchers would be able to troubleshoot by spotting outliers. For instance, we thought that researchers could doubt the quality of a recording when the child had an unexpectedly low number of vocalizations (e.g. if the child's vocalizations were systematically missed due to the recorder not being worn by the child, or the child's voice deviating from the training data) or high (e.g. if siblings were present and their vocalizations were incorrectly attributed to the key child). However, we have not been able to prove this with the data available to us. We suspect this would require the development of norms, a costly process that involves recording a representative sample of participants to determine what number of vocalizations for each class are typical, above average, or below average for that group. 

\section{Discussion}

Long-form recordings aspire to give us a truthful view of the speech spontaneously produced around and by young children. In this paper, we provide readers with an entry point to this technique and the many resources that are accumulating around it. Long-form recording data are not a silver bullet. Fifteen years of experience suggest there could be a wide array of challenges to metrics emerging from automated analyses, which we spelled out in Table \ref{tab:problem_classification}. Although automated metrics of audio quality are promising indices of algorithm performance, we think that additional work is needed to find automated ways of troubleshooting all such issues, since our analyses of private datasets suggest that some effects are not obvious (Sections 5.2.2 and 5.2.3). We hope future work further explores automated indices of data quality that may threaten our interpretation of metrics based on automated analyses of long-form recordings, particularly as current devices are used in more diverse environments.

\

\printendnotes[interspeech]

\bibliographystyle{IEEEtran}
\bibliography{mybib}

\end{document}